\documentclass[aps,pra,twocolumn,superscriptaddress,nobalancelastpage]{revtex4-1}
\usepackage{amsmath,amssymb,mathtools}
\usepackage[dvipdfmx]{graphicx}
\usepackage[dvipdfmx,colorlinks=true,bookmarks=true,citecolor=blue,linkcolor=blue,urlcolor=blue,breaklinks=true]{hyperref}

\graphicspath{{fig/}}
\usepackage{dcolumn}
\usepackage{bm}
\usepackage[T1]{fontenc}
\usepackage{textcomp}
\usepackage{color}
\usepackage{mathrsfs}

\begin{document}

\title{Direct and alternating magnon spin currents across a junction interface   \\
irradiated by linearly polarized laser}

\author{Kouki Nakata}
\affiliation{Advanced Science Research Center, Japan Atomic Energy Agency, Tokai, Ibaraki 319-1195, Japan}

\author{Yuichi Ohnuma}
\affiliation{Research Center for Advanced Science and Technology (RCAST),
The University of Tokyo, Meguro, Tokyo 153-8904, Japan
}

\date{\today}

\begin{abstract}
The developments in the field of quantum optics raise expectations that  laser-matter coupling is a promising building block for magnonics. Here, we propose a method for the generation of direct and alternating spin currents of magnons across the junction interface irradiated by linearly polarized laser. In a junction of ferromagnetic insulators with a large electronic gap, the spin angular momentum is exchanged during the tunneling process of magnons across the junction interface. The advanced technology in the field of plasmonics and metamaterials realizes that spins irradiated by the laser field interact only with the magnetic component of the laser through the Zeeman coupling. Using an analytic perturbation theory, we provide a general formula for magnon transport induced by the inversion symmetry breaking across the junction interface. Then, we show that those spin currents are enhanced by the ferromagnetic resonance, and the period of the ac spin current is one-half of that of the laser magnetic field. Finally, we estimate the magnitude of the spin current, and find that it will be within experimental reach.
\end{abstract}

\maketitle

\section{Introduction}
\label{sec:intro}

Toward efficient transmission of information 
that goes beyond what is offered by conventional electronics,
the last decade has seen a rapid development of magnon-based spintronics, dubbed magnonics~\cite{MagnonSpintronics,ReviewMagnon},
aiming at utilizing the quantized spin waves, magnons,
as a carrier of information in units of the Bohr magneton $\mu_{\text{B}}$.
A most promising strategy for this goal is to use magnetic insulators,
which are free from drawbacks of conventional electronics 
such as significant energy loss due to Joule heating.
Thus, inventing methods to handle a flexible manipulation 
of magnon transport in magnetic insulators,
in the complete absence of any conducting metallic elements,
is a central task in the field of magnonics.

The recent developments in the field of
plasmonics~\cite{Ciappina2017RepProgPhys}
and metamaterials~\cite{Mukai2014APL} 
raise expectations that 
laser-matter coupling~\cite{Mukai2014APL,LaserPhotoExp,Ciappina2017RepProgPhys,LaserPhotoExp3}
is an important building block for this holy grail of magnonics.
Using the optical method~\cite{Kirilyuk,KimelMagnetizationReversalExp4,KimelNatureIFaraday,Kimel7,Kimel8,AFopticsReview}, 
the reversal of magnetization was achieved 
experimentally~\cite{OtherOpticalBarnett,KimelMagnetizationReversalExp2,KimelMagnetizationReversalExp3,KimelMagnetizationReversalExp5}.
An optical analog of the conventional Barnett effect~\cite{Barnett,Barnett2,ReviewSpinMechatronics},
dubbed the optical Barnett effect~\cite{OtherOpticalBarnett2,OtherOpticalBarnett3}
(i.e., laser-induced magnetization~\cite{FloquetST2,FloquetST})~\footnote{
See Refs.~\cite{FloquetST2,FloquetST} 
for the importance of modulating laser frequency adiabatically
by the chirping technique~\cite{chirping,chirping2}.
See also Ref.~\cite{KNST_OBarnett}
for the difference between the optical Barnett effect
and the inverse Faraday effect~\cite{Kirilyuk,KimelNatureIFaraday}.},
and the resulting magnon Josephson effect~\cite{OMJ}
were proposed theoretically
by means of circularly polarized laser.

In this paper, using linearly polarized laser,
we propose a method for the generation of 
direct (dc) and alternating (ac) spin currents of magnons
across a junction interface 
due to the inversion symmetry breaking.
We consider a junction formed by ferromagnetic insulators 
with a large electronic gap,
where the spin angular momentum is exchanged 
during the tunneling process of magnons across the junction interface.
In addition to the large electronic gap,
owing to the state-of-the-art technology 
in the field of plasmonics and metamaterials
that realizes strong magnetic field pulses with small electric field~\cite{Ciappina2017RepProgPhys,Mukai2014APL},
spins in the laser field interact 
only with the magnetic component of the laser 
through the Zeeman coupling.
By means of an analytic perturbation theory,
we provide a general formula for magnon transport 
induced by the inversion symmetry breaking 
across the junction interface.
Then, we find that
the ferromagnetic resonance enhances those spin currents,
and the period of the ac spin current is 
one-half of that of the laser magnetic field.
Finally, we estimate the magnitude of the spin current,
and show that it will be within experimental reach.

This paper is organized as follows.
We explain the setup of our study
in Sec.~\ref{sec:system},
and investigate magnon transport across the junction interface
in Sec.~\ref{sec:SHG}.
Then, 
estimating the values of the magnon spin current,
we propose a method for observation,
and provide a few remarks in Sec.~\ref{sec:discussion}.
Finally, we summarize in Sec.~\ref{sec:conclusion}.
Technical details are described in the Appendix.

\begin{figure}[t]
\begin{center}
\includegraphics[width=9cm,clip]{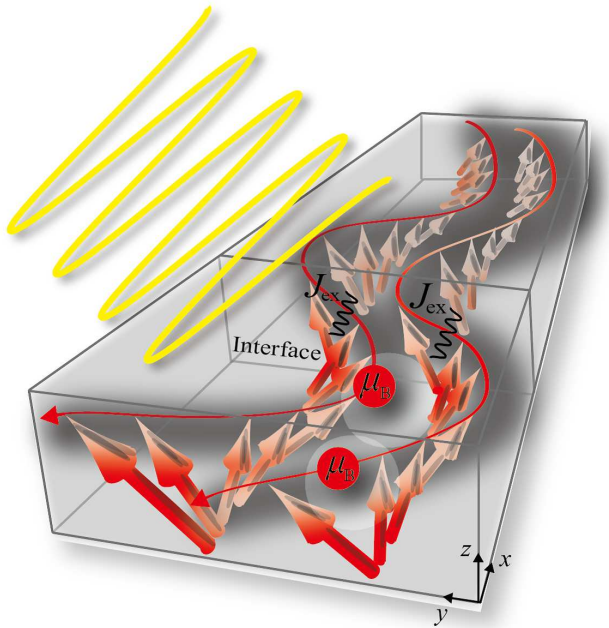}
\caption{Schematic picture of the junction.
The junction consists of
three-dimensional ferromagnetic insulators
where localized spins are polarized along the $z$ direction.
The two ferromagnetic insulators are separated 
by a thin film of a nonmagnetic insulator 
and the boundary spins in the left and the right ferromagnetic insulators 
are weakly exchange coupled, 
which results in the tunneling process of magnons 
with the amplitude $J_{\text{ex}} $
across the junction interface.
The linearly polarized laser, 
i.e., time-periodic transverse magnetic field,
with the field strength of being spatially uniform 
is applied to the two ferromagnetic insulators.
}
\label{fig:SHG}
\end{center}
\end{figure}

\section{System}
\label{sec:system}

We consider a junction of 
three-dimensional ferromagnetic insulators
with a large electronic gap.
The two ferromagnetic insulators are aligned along the $x$ direction
and the localized spins are polarized along the $z$ direction,
see Fig.~\ref{fig:SHG}~\cite{magnonWF}.
A finite overlap of the wave function of the boundary spins
in the left and the right ferromagnetic insulators
results in an exchange interaction,
which induces a tunneling process of magnons
across the junction interface.
During the tunneling process,
the spin angular momentum is exchanged 
and it may be described in general
by the Hamiltonian~\cite{magnonWF,KNmagnonNoiseJunction}
$ {\mathcal{H}}_{\text{ex}} 
= - J_{\text{ex}} 
\sum_{{\mathbf{k}}_{\perp }} 
\sum_{k_x, k_x^{\prime}}
a_{{\text{L}}, {\mathbf{k}}} 
a^{\dagger }_{{\text{R}}, {\mathbf{k}}^{\prime}} 
+ {\text{H. c.}}$,
where
the bosonic operator
$  a^{(\dagger)}_{\text{L/R}} $
annihilates (creates) magnons 
at the boundary of the left/right insulator 
with the wave numbers
${\mathbf{k}}\coloneqq(k_x, k_y, k_z)$,
${\mathbf{k}}^{\prime}\coloneqq(k_x^{\prime}, k_y, k_z)$, 
and 
${\mathbf{k}}_{\perp }\coloneqq(0, k_y, k_z)  $,
and where
$J_{\text{ex}} >0$ represents the tunneling amplitude of magnons
across the junction interface
and 
the value depends on the spin quantum number of the boundary spins 
in each insulator.
The ferromagnetic insulators
are separated by a thin film of a nonmagnetic insulator, 
and the boundary spins in the left and the right ferromagnetic insulators
are weakly exchange coupled in that 
the magnitude of $J_{\text{ex}}$
is small compared with the spin exchange interaction 
between the nearest-neighbor spins in each insulator.
See Refs.~\cite{magnonWF,KNmagnonNoiseJunction}
for the details of the relation 
between the magnon Hamiltonian
$ {\mathcal{H}}_{\text{ex}} $
and the microscopic spin model
through a Holstein-Primakoff expansion to leading order~\cite{HP}.

Then, we apply the linearly polarized laser of the frequency $\Omega$ 
to the junction of the two ferromagnetic insulators
with a large electronic gap.
Due to the large electronic gap,
spins in the laser 
interact only with the magnetic component of the laser field
through the Zeeman coupling.
Since we assume that the magnetic field amplitude
is spatially uniform,
the laser field (i.e., time-periodic transverse magnetic field)
excites the zero mode of magnons in each insulator~\footnote{Note that in Ref.~\cite{KPD}, the laser field is applied only to one side.}
and the process in the left (right) insulator
is described~\cite{KPD}
by the time-periodic Hamiltonian $V_{\text{L(R)}} (t)$:
\begin{subequations}
\begin{align}
V_{\text{L}} (t)
&=  
 \Gamma_{\text{L}}(t)
(a_{{\text{L}},  {\mathbf{k}}=0} 
+ a^{\dagger }_{{\text{L}}, {\mathbf{k}}=0}),
 \label{eq:VL}  \\
V_{\text{R}} (t)
 &=  
 \Gamma_{\text{R}}(t)
(a_{{\text{R}},  {\mathbf{k}}=0} 
+ a^{\dagger }_{{\text{R}}, {\mathbf{k}}=0}), 
\label{eq:VR}
\end{align}
\end{subequations}
with
\begin{subequations}
\begin{align}
 \Gamma_{\text{L}}(t)
 &=2\Gamma_{0({\text{L}})} {{\text{cos}}(\Omega t)}, 
 \label{eq:GammaL} \\
 \Gamma_{\text{R}}(t)
 &=2\Gamma_{0({\text{R}})} {{\text{cos}}(\Omega t)},
  \label{eq:GammaR}
\end{align}
\end{subequations}
where 
$ 2\Gamma_{0({\text{L/R}})} $
represents the magnitude of the interaction 
between the laser magnetic field and magnons in the left/right insulator.
Since the magnitude depends on the spin quantum number,
it is 
$ \Gamma_{0(\text{L})}  \neq \Gamma_{0(\text{R})}  $
in general.
For simplicity,
assuming that the $g$-factor of spins in the left insulator 
is identical to that of the right one,
we denote it as $g$.
Since we apply a weak laser field 
in that the magnitude of $ 2\Gamma_{0({\text{L/R}})} $
is small compared with the spin exchange interaction 
between the nearest-neighbor spins in each insulator,
the description by means of magnon degrees of freedom 
through the Holstein-Primakoff expansion to leading order
[Eqs.~\eqref{eq:VL} and~\eqref{eq:VR}]
remains valid
and the laser magnetic field $V_{\text{L(R)}} (t) $
can be treated as a perturbation term.

We remark that since the laser excites the zero mode of magnons,
the result does not depend on the details of the energy dispersion relation, 
e.g., the spin stiffness constant,
except the magnon energy gap
$\Delta_{\text{L(R)}}$
for the zero mode
in the left (right) insulator.
The magnon energy gap is formed by, 
e.g., a spin anisotropy~\cite{KSJD}.
Later, for convenience,
we phenomenologically introduce
the lifetime~\cite{QBEmagnon}
of magnons $\tau_{\text{L(R)}}$ 
in the left (right) insulator
through the Green's function~\cite{tatara}.

\section{Magnon transport in junction}
\label{sec:SHG}

\subsection{Spin currents of magnons}
\label{subsec:SHG1}

Through the Heisenberg equation of motion,
the expectation value of the magnon spin current
across the junction interface
$\langle I (t) \rangle  $,
from the right to the left insulator,
is given as~\cite{SeeAppendixOnly}
\begin{equation}
\langle I (t) \rangle 
= 
2 g \mu _{\text{B}} \frac{J_{\text{ex}}}{\hbar} 
{\text{Im}} \langle {\mathcal{A}}(t)  \rangle
+O({J_{\text{ex}}}^2),
\label{eq:I}
\end{equation}
where $\hbar$ denotes the reduced Planck constant and 
\begin{equation}
 {\mathcal{A}}(t)
\coloneqq
\sum_{{\mathbf{k}}, k_x^{\prime}}
a_{{\text{L}},{\mathbf{k}}}(t)  
a^{\dagger }_{{\text{R}},{\mathbf{k}}^{\prime}}(t).
\label{eq:A}
\end{equation}
Thus, first, we calculate 
$  \langle {\mathcal{A}}(t)  \rangle 
= O({J_{\text{ex}}}^0 {\Gamma_{0({\text{L/R}})}}^2)$
using a perturbation theory.
Then, taking the imaginary component of
$  \langle {\mathcal{A}}(t)  \rangle$,
we evaluate the magnon spin current 
$ \langle I (t) \rangle  
= O(J_{\text{ex}} {\Gamma_{0({\text{L/R}})}}^2)$
in the junction subjected to the laser magnetic field.

A straightforward perturbative calculation in $ \Gamma_{0({\text{L/R}})}$
based on the Schwinger-Keldysh formalism~\cite{Schwinger,Schwinger2,Keldysh},
and up to $ {\Gamma_{0({\text{L/R}})}}^2$,
provides~\cite{SeeAppendixOnly}
\begin{align}
\langle {\mathcal{A}}(t) \rangle
=\frac{1}{\hbar^2} 
& \int_{{\text{c}}} d \tau_1
\int_{{\text{c}}} d \tau_2
\Gamma_{\text{L}}(\tau_1)
\Gamma_{\text{R}}(\tau_2)  
\nonumber  \\
&\cdot
G_{{\text{L}}, {\mathbf{k}}=0}(\tau, \tau_1)
G_{{\text{R}}, {\mathbf{k}}=0}(\tau_2, \tau)|_{\tau \rightarrow t},
\label{eq:A1}
\end{align}
where
$ G_{{\text{L(R)}}} $
is the 
contour-ordered Green's function~\cite{haug,kita,tatara}
of magnons in the left (right) insulator,
and 
$\tau$ and $\tau_{1(2)}$
are the contour variables 
defined on the Schwinger-Keldysh closed time path~\cite{rammer}
$ {\text{c}} $,
$ {\text{c}}\coloneqq{\text{c}}_{\rightarrow}+{\text{c}}_{\leftarrow} $,
consisting of
the forward path
$ {\text{c}}_{\rightarrow} $
and
the backward path
$ {\text{c}}_{\leftarrow}  $.
Here, for convenience, we take $\tau$ on the forward path.
Even when $\tau$ is located on the backward path, 
the result remains unchanged.
Using the Langreth theorem~\cite{LangrethTheorem},
it becomes
\begin{align}
\langle {\mathcal{A}}(t) \rangle
= \frac{1}{\hbar^2} 
&\int_{-\infty}^{\infty}dt_1
\int_{-\infty}^{\infty}dt_2
\Gamma_{\text{L}}(t_1)
\Gamma_{\text{R}}(t_2)
\nonumber \\
&\cdot
[G^{\text{t}}_{{\text{L}},  {\mathbf{k}}=0}(t, t_1)
G^{\text{t}}_{{\text{R}},  {\mathbf{k}}=0}(t_2, t)  
\nonumber  \\
&-G^{\text{t}}_{{\text{L}},  {\mathbf{k}}=0}(t, t_1)
G^{\text{>}}_{{\text{R}},  {\mathbf{k}}=0}(t_2, t) 
\nonumber  \\
&-G^{\text{<}}_{{\text{L}},  {\mathbf{k}}=0}(t, t_1)
G^{\text{t}}_{{\text{R}},  {\mathbf{k}}=0}(t_2, t) 
\nonumber  \\
&+G^{\text{<}}_{{\text{L}},  {\mathbf{k}}=0}(t, t_1)
G^{\text{>}}_{{\text{R}},  {\mathbf{k}}=0}(t_2, t)],
\label{eq:A2}
\end{align}
where
$G^{\text{t}} $,
$G^{<}$,
and 
$G^{>}$
are
the time-ordered,
lesser,
and greater Green's functions, respectively.
Through the identities,
$ G^{\text{t}}_{{\mathbf{k}}}(t, t_1)  
= G^{\text{r}}_{{\mathbf{k}}}(t, t_1) 
+ G^{<}_{{\mathbf{k}}}(t, t_1) $
and
$ G^{<}_{{\mathbf{k}}}(t, t_1)  
- G^{>}_{{\mathbf{k}}}(t, t_1) 
= G^{\text{a}}_{{\mathbf{k}}}(t, t_1) 
- G^{\text{r}}_{{\mathbf{k}}}(t, t_1) $,
it reduces to
$  \langle {\mathcal{A}}(t) \rangle
= 
(1/\hbar^2)
\int_{-\infty}^{\infty}dt_1
\int_{-\infty}^{\infty}dt_2
\Gamma_{\text{L}}(t_1)
\Gamma_{\text{R}}(t_2)  
G^{\text{r}}_{{\text{L}}}(t, t_1)
G^{\text{a}}_{{\text{R}}}(t_2, t)  $,
where
$ G^{\text{r(a)}} $
is the retarded (advanced) Green's function.
After the Fourier transform of the Green's functions,
we perform the time integral 
and the resulting Dirac delta function for the frequency.
Finally, we obtain
\begin{align}
 \langle {\mathcal{A}}(t) \rangle
=
\Gamma_{0(\text{L})}
\Gamma_{0(\text{R})}
&[
(G^{\text{r}}_{{\text{L}},  {\mathbf{k}}=0, -\Omega}
G^{\text{a}}_{{\text{R}},  {\mathbf{k}}=0, -\Omega} \nonumber  \\
&+
G^{\text{r}}_{{\text{L}},  {\mathbf{k}}=0, \Omega}
G^{\text{a}}_{{\text{R}},  {\mathbf{k}}=0, \Omega}) \nonumber  \\
&+
G^{\text{r}}_{{\text{L}},  {\mathbf{k}}=0, -\Omega}
G^{\text{a}}_{{\text{R}},  {\mathbf{k}}=0, \Omega}
{\text{e}}^{2i\Omega t}  \nonumber \\
&+
G^{\text{r}}_{{\text{L}},  {\mathbf{k}}=0, \Omega}
G^{\text{a}}_{{\text{R}},  {\mathbf{k}}=0, -\Omega}
{\text{e}}^{-2i\Omega t} 
].
\label{eq:A3}
\end{align}
Thus, we find from Eqs.~\eqref{eq:A3} and~\eqref{eq:I}
that the tunneling spin current of magnons 
is described essentially as the product of the magnetic susceptibility for each insulator.
Those are summarized as follows:
\begin{equation}
 \langle {\mathcal{A}}(t) \rangle
=
\Gamma_{0(\text{L})}
\Gamma_{0(\text{R})}
\sum_{n=0,\pm2} 
A_n {\text{e}}^{ni\Omega t} ,
\label{eq:An}
\end{equation}
with the coefficients
\begin{subequations}
\begin{align}
A_2
&=
G^{\text{r}}_{{\text{L}},  {\mathbf{k}}=0, -\Omega}
G^{\text{a}}_{{\text{R}},  {\mathbf{k}}=0, \Omega}, 
\label{eq:A2}  \\
A_{-2}
&=
G^{\text{r}}_{{\text{L}},  {\mathbf{k}}=0, \Omega}
G^{\text{a}}_{{\text{R}},  {\mathbf{k}}=0, -\Omega}, 
\label{eq:A-2} \\
A_0
&=
G^{\text{r}}_{{\text{L}},  {\mathbf{k}}=0, -\Omega}
G^{\text{a}}_{{\text{R}},  {\mathbf{k}}=0, -\Omega}
+
G^{\text{r}}_{{\text{L}},  {\mathbf{k}}=0, \Omega}
G^{\text{a}}_{{\text{R}},  {\mathbf{k}}=0, \Omega}.
\label{eq:A0}
\end{align}
\end{subequations}

\subsection{dc and ac components}
\label{subsec:SHG2}

We evaluate 
$  {\text{Im}} {\langle {\mathcal{A}}(t) \rangle}  $
and provide an analytical formula for the magnon spin current 
across the junction interface
[Eq.~\eqref{eq:I}].
To this end,
defining 
\begin{equation}
A_{\Omega} \coloneqq A_2 \in {\mathbb{C}},
\label{eq:AOmega}
\end{equation}
we introduce the real variables
$ A_{\Omega}^{\text{R}} \in {\mathbb{R}} $
and
$ A_{\Omega}^{\text{I}} \in {\mathbb{R}} $
as
$A_{\Omega} 
=A_{\Omega}^{\text{R}}+i A_{\Omega}^{\text{I}}
\in {\mathbb{C}} $
for convenience,
and describe the magnon spin current
in terms of 
$ A_{\Omega}^{\text{R}}= {\text{Re}}(A_2) $
and
$ A_{\Omega}^{\text{I}}= {\text{Im}}(A_2)$.
Using the identity
\begin{equation}
A_{-\Omega} 
=A_{-2},
\label{eq:AminusOmega}
\end{equation}
a straightforward calculation 
starting from Eq.~\eqref{eq:An} provides
\begin{align}
 {\text{Im}} {\langle {\mathcal{A}}(t) \rangle}
=&
\Gamma_{0(\text{L})}
\Gamma_{0(\text{R})}
[
(A_{\Omega}^{\text{I}}+A_{-\Omega}^{\text{I}})
{\text{cos}}({2\Omega t})  
\nonumber  \\
&+(A_{\Omega}^{\text{R}}-A_{-\Omega}^{\text{R}})
{\text{sin}}({2\Omega t})
+{\text{Im}}(A_0)].
\label{eq:ImA}
\end{align}
We find that the magnon spin current [Eq.~\eqref{eq:I}]
consists of two parts:
The dc component, $ {\text{Im}}(A_0)  $,
and the ac one.
Finally,
we obtain the magnon spin current across the junction interface as
\begin{equation}
\langle I (t) \rangle 
= \langle I_{\text{{ac}}} (t) \rangle 
+ \langle I_{\text{dc}}  \rangle 
+O({J_{\text{ex}}}^2),
\label{eq:IshgDC}
\end{equation}
where
the ac component 
$ \langle I_{\text{{ac}}} (t) \rangle 
= O(J_{\text{ex}} \Gamma_{0(\text{L})}\Gamma_{0(\text{R})})  $
and the dc one
$  \langle I_{\text{dc}}  \rangle 
= O(J_{\text{ex}} \Gamma_{0(\text{L})}\Gamma_{0(\text{R})})  $
are described as
\begin{subequations}
\begin{align}
\langle I_{\text{{ac}}} (t) \rangle 
\coloneqq& 
2 g \mu _{\text{B}} \frac{J_{\text{ex}}}{\hbar} 
\sqrt{(A_{\Omega}^{\text{I}}+A_{-\Omega}^{\text{I}})^2
+(A_{\Omega}^{\text{R}}-A_{-\Omega}^{\text{R}})^2} 
\nonumber  \\
&\cdot
\Gamma_{0(\text{L})}\Gamma_{0(\text{R})} 
{\text{cos}}({2\Omega t}+\theta_0), 
\label{eq:SHG}  \\
 \langle I_{\text{dc}}  \rangle 
\coloneqq&
2 g \mu _{\text{B}} \frac{J_{\text{ex}}}{\hbar}  
\Gamma_{0(\text{L})}\Gamma_{0(\text{R})}
{\text{Im}}(A_0),  
\label{eq:dc}
\end{align}
\end{subequations}
with the phase $\theta_0 $ 
characterized as
\begin{subequations}
\begin{align}
{\text{sin}}\theta_0
&=
-\frac{A_{\Omega}^{\text{R}}-A_{-\Omega}^{\text{R}}}{\sqrt{(A_{\Omega}^{\text{I}}+A_{-\Omega}^{\text{I}})^2
+(A_{\Omega}^{\text{R}}-A_{-\Omega}^{\text{R}})^2}},
\label{eq:sin}  \\
{\text{cos}}\theta_0
&=
\frac{A_{\Omega}^{\text{I}}+A_{-\Omega}^{\text{I}}}{\sqrt{(A_{\Omega}^{\text{I}}+A_{-\Omega}^{\text{I}})^2
+(A_{\Omega}^{\text{R}}-A_{-\Omega}^{\text{R}})^2}}.
\label{eq:cos} 
\end{align}
\end{subequations}
This is the general formula for magnon transport
in the junction irradiated by the laser field:
The main result of this paper.
For convenience, introducing the variables 
$C_{\text{I}} :=  A_{\Omega}^{\text{I}} + A_{-\Omega}^{\text{I}}  $
and
$C_{\text{R}} :=  A_{\Omega}^{\text{R}} - A_{-\Omega}^{\text{R}}  $,
the phase of the ac component $ \theta_0$
is characterized, as an example, as follows:
\begin{subequations}
\begin{align}
\theta_0&=0 
\  \  \  \  \   \   \   \text{for}   \
C_{\text{I}} >0 \   \text{and}   \
C_{\text{R}} =0.  
\label{eq:theta01} \\
\theta_0&=-\pi/2 \  \text{for}   \
C_{\text{I}} =0 \   \text{and}   \
C_{\text{R}} >0. 
\label{eq:theta02}  \\
\theta_0&=-\pi/4 \  \text{for}   \
C_{\text{I}}=C_{\text{R}} >0.  
\label{eq:theta03}  \\
\theta_0&=\pi/4  
\ \  \   \text{for}   \
C_{\text{I}}=-C_{\text{R}} >0.
\label{eq:theta04} 
\end{align}
\end{subequations}

We find from Eq.~\eqref{eq:SHG} that 
the ac spin current of the period $\pi/\Omega$,
\begin{equation}
 \langle I_{\text{{ac}}} (t) \rangle 
= \langle I_{\text{{ac}}} (t+\pi/\Omega) \rangle ,
\label{eq:SHGformula}
\end{equation}
arises from the laser magnetic field 
of the period $2\pi/\Omega$
[Eqs.~\eqref{eq:GammaL} and~\eqref{eq:GammaR}]
\begin{equation}
\Gamma_{\text{L(R)}}(t) = \Gamma_{\text{L(R)}}(t+2\pi/\Omega).
\label{eq:LaserPeriod}
\end{equation}
The period of the ac spin current is one-half of that of the laser magnetic field.
We remark that this spin current of magnons is 
analogous to the Josephson spin current of magnons~\cite{KKPD,OMJ}
in that those spin currents across the junction interface
arise as the process of $O(J_{\text{ex}})$
in $ J_{\text{ex}} $.
This is in contrast to the spin current of thermal magnons, 
which arises as the process of $O({J_{\text{ex}}}^2)$ 
in $ J_{\text{ex}} $,
induced by the applied temperature difference~\cite{magnonWF}.

\subsection{Inversion symmetry breaking}
\label{subsec:SHG3}

From a practical viewpoint,
we describe $ A_{\Omega}^{\text{R(I)}}$ and $A_0$
in terms of the magnon lifetime 
by phenomenologically introducing it~\footnote{
Here, we phenomenologically assume that 
the finite lifetime of magnons is caused by impurity scattering~\cite{QBEmagnon}.}
through the Green's function~\cite{tatara},
e.g.,
$ G^{\text{r}}_{{\text{L}},  {\mathbf{k}}=0, \Omega}
=1/[\hbar \Omega - \Delta_{\text{L}}+i\hbar/(2\tau_{\text{L}})]
=(G^{\text{a}}_{{\text{L}},  {\mathbf{k}}=0, \Omega})^{*}$,
where
$ \tau_{\text{L(R)}} $ and $\Delta_{\text{L(R)}}$
are the lifetime and the energy gap of magnons
for the zero mode
in the left (right) insulator, respectively,
and
$(G^{\text{a}}_{{\text{L}},  {\mathbf{k}}=0, \Omega})^{*}$
represents the complex conjugate of
$G^{\text{a}}_{{\text{L}},  {\mathbf{k}}=0, \Omega}$.
Then, those are characterized as~\cite{SeeAppendixOnly}
\begin{subequations}
\begin{align}
A_{\Omega}^{\text{R}}
=&
-\frac{(\hbar \Omega + \Delta_{\text{L}})(\hbar \Omega - \Delta_{\text{R}})-(\hbar/2\tau_{\text{L}})(\hbar/2\tau_{\text{R}})}{[(\hbar \Omega + \Delta_{\text{L}})^2+(\hbar/2\tau_{\text{L}})^2][(\hbar \Omega - \Delta_{\text{R}})^2+(\hbar/2\tau_{\text{R}})^2]},
\label{eq:AOmegaR}  \\
A_{\Omega}^{\text{I}}
=&
-\frac{(\hbar/2\tau_{\text{L}})(\hbar \Omega - \Delta_{\text{R}})+(\hbar/2\tau_{\text{R}})(\hbar \Omega + \Delta_{\text{L}})}{[(\hbar \Omega + \Delta_{\text{L}})^2+(\hbar/2\tau_{\text{L}})^2][(\hbar \Omega - \Delta_{\text{R}})^2+(\hbar/2\tau_{\text{R}})^2]}, 
\label{eq:AOmegaI}  \\
{\text{Im}}(A_0)
&=
\frac{(\hbar/2\tau_{\text{L}})(\hbar \Omega + \Delta_{\text{R}})
-(\hbar/2\tau_{\text{R}})(\hbar \Omega + \Delta_{\text{L}})}{[(\hbar \Omega + \Delta_{\text{L}})^2+(\hbar/2\tau_{\text{L}})^2][(\hbar \Omega + \Delta_{\text{R}})^2+(\hbar/2\tau_{\text{R}})^2]}  
\nonumber  \\
&+
\frac{(\hbar/2\tau_{\text{L}})(-\hbar \Omega + \Delta_{\text{R}})
-(\hbar/2\tau_{\text{R}})(-\hbar \Omega + \Delta_{\text{L}})}{[(\hbar \Omega - \Delta_{\text{L}})^2+(\hbar/2\tau_{\text{L}})^2][(\hbar \Omega - \Delta_{\text{R}})^2+(\hbar/2\tau_{\text{R}})^2]}.
\label{eq:ImA0}
\end{align}
\end{subequations}
These show that tuning the laser frequency,
\begin{equation}
\hbar \Omega = \Delta_{\text{L(R)}},
\label{eq:resonance}
\end{equation}
the ferromagnetic resonance enhances
those magnon spin currents across the junction interface.

We remark that 
when the left insulator is identical to the right one,
i.e.,
$ \Delta_{\text{L}}= \Delta_{\text{R}} $
and
$ \tau_{\text{L}}= \tau_{\text{R}} $,
the coefficients become
\begin{subequations}
\begin{align}
A_{\Omega}^{\text{I}}+A_{-\Omega}^{\text{I}}
&=0,  
\label{eq:AOmegaILR} \\
A_{\Omega}^{\text{R}}-A_{-\Omega}^{\text{R}}
&=0, 
\label{eq:AOmegaRLR} \\
{\text{Im}}(A_0)
&=0.
\label{eq:ImA00}
\end{align}
\end{subequations}
This results in
\begin{subequations}
\begin{align}
\langle I_{\text{{ac}}} (t) \rangle 
&=0, 
\label{eq:Ishg0} \\
 \langle I_{\text{dc}}  \rangle 
&=0,  
\label{eq:Idc0} \\
\langle I (t) \rangle  
&=0,
\label{eq:I000}
\end{align}
\end{subequations}
and there are no magnon spin currents across the junction interface.
This shows that the dc component of the magnon spin current
arises from the difference of the magnon energy gap
in the junction subjected to the laser field: 
In the junction out of equilibrium, the magnon energy gap works 
as a nonequilibrium spin chemical potential~\cite{Basso2,MagnonChemicalWees,MagnonG,YacobyChemical,demokritov}
(i.e., a potential in an effective magnetic field~\cite{SilsbeeMagnetization})
and the difference induces magnon transport across the junction interface~\cite{Haldane2,magnon2,Fujimoto,KSJD,magnonWF,KNmagnonNoiseJunction}.
When
$ \tau_{\text{L}}= \tau_{\text{R}} =:\tau$
while
$ \Delta_{\text{R}} \neq \Delta_{\text{L}} $,
the inversion symmetry across the junction interface violates.
Then, the dc component is proportional to
the difference of the magnon energy gap,
$ \Delta_{\text{R}} -  \Delta_{\text{L}} $,
as follows:
\begin{align}
\langle I_{\text{dc}}  \rangle
&=2g\mu_{\text{B}} \frac{J_{\text{ex}}}{\hbar}
\Gamma_{0(\text{L})}
\Gamma_{0(\text{R})}
\frac{\hbar}{2\tau} (\Delta_{\text{R}} -  \Delta_{\text{L}})   \\
&\times
\Bigg[
\frac{1}{[(\hbar \Omega+\Delta_{\text{L}})^2+(\hbar/2\tau)^2]
[(\hbar \Omega+\Delta_{\text{R}})^2+(\hbar/2\tau)^2]}  
\nonumber \\
&+
\frac{1}{[(\hbar \Omega-\Delta_{\text{L}})^2+(\hbar/2\tau)^2]
[(\hbar \Omega-\Delta_{\text{R}})^2+(\hbar/2\tau)^2]}
\Bigg].  
\nonumber
\label{eq:IdcDelta}
\end{align}
Hence, when $ \Delta_{\text{R}} > \Delta_{\text{L}} $,
the dc spin current flows from the right to the left insulator
due to the inversion symmetry breaking.

\section{Discussion}
\label{sec:discussion}

For an estimate,
we assume the following parameter values~\cite{GilbertInsulator,LLGspintroReview,YIGBEC2008}:
$\Delta_{\text{L}}=\hbar \Omega=5 \  \mu$eV,
$\Delta_{\text{R}}=6 \  \mu$eV,
$\Gamma_{0(\text{L})} = \Gamma_{0(\text{R})}=0.5  \  \mu$eV,
$\tau=10  \   \mu$s~\cite{GilbertInsulator,LLGspintroReview,QBEmagnon},
$J_{\text{ex}}=35  \  \mu$eV~\cite{YIGBEC2008}.
Then, following Ref.~\cite{Kishine2},
we estimate the dc spin current $\langle I_{\text{dc}} \rangle $
in electric units $e/g\mu_{\text{B}}$,
and find that it amounts to
$\langle I_{\text{dc}} \rangle e/g\mu_{\text{B}} 
\sim 0.05$ mA.
We believe, while being small 
compared with the one ($0.1$ mA)~\cite{Kishine2} 
for antiferromagnets,
still it will be within experimental reach.
We expect, to the best of our knowledge, 
that to use the inverse spin Hall effect~\cite{ISHE1}
by attaching a metal to the insulating magnet
will be one of the most promising strategies for observation.

In this paper focusing on the junction
and assuming the weak exchange coupling
between the left and right ferromagnetic insulator,
we have studied the tunneling spin current of magnons
across the junction interface
irradiated by linearly polarized laser.
It will be of significance to develop this work
into bulk insulators
by taking into account higher-order terms
in the Holstein-Primakoff expansion
and thus studying the nonlinear terms
in the magnetic susceptibility.
We leave the advanced study for future work.
We remark that 
Ref.~\cite{Kishine2}
proposed the optically induced spin current of magnons
in the bulk of antiferromagnetic insulators
irradiated by a circularly polarized electromagnetic field,
where, analogous to our present work,
the field-induced spin current is described 
by the product of the susceptibility,
and the current is enhanced by 
(antiferromagnetic) resonance.

\section{Conclusion}
\label{sec:conclusion}

Using the junction of the ferromagnetic insulators
irradiated by linearly polarized laser,
we have proposed the method for the generation of 
dc and ac spin currents of magnons across the junction interface,
and provided the analytical formula for magnon transport
induced by the inversion symmetry breaking.
We have shown that the applied laser field excites the zero mode of magnons, and tuning the laser frequency to the magnon energy gap for the zero mode, the ferromagnetic resonance enhances those spin currents across the junction interface.
The period of the ac spin current is one-half of that of the laser magnetic field.
We hope that our proposal serves as a key ingredient 
for magnonics.

\acknowledgements

We would like to thank S. Takayoshi
for fruitful discussions and feedback on this manuscript.
The author (K. N.) is grateful also to
S. Uchino and Y. Araki for useful discussions,
and H. Chudo for helpful feedback on the experimental feasibility.
We also sincerely thank anonymous referees 
for the valuable comments 
that helped us drastically improve the manuscript.
We acknowledge support
by JSPS KAKENHI Grant Number JP20K14420 (K. N.)
and JP22K03519 (K. N.),
by Leading Initiative for Excellent Young Researchers, 
MEXT, Japan (K. N.),
and by JST ERATO Grant No. JPMJER1601 (Y. O.).

\begin{widetext}

\appendix*

\section{Magnon transport across the junction interface}
\label{sec:Appendix}

In this Appendix,
we provide some details of the straightforward calculation
for the magnon spin current across the junction interface
$ \langle I (t) \rangle 
= O(J_{\text{ex}} {\Gamma_{0({\text{L/R}})}}^2) $,
\begin{equation}
\langle I (t) \rangle 
= 
2 g \mu _{\text{B}} \frac{J_{\text{ex}}}{\hbar} 
{\text{Im}} \langle {\mathcal{A}}(t) \rangle 
+O({J_{\text{ex}}}^2),
\label{eq:Appendix0}
\end{equation}
by evaluating
$ \langle {\mathcal{A}}(t) \rangle  
= O({J_{\text{ex}}}^0 {\Gamma_{0({\text{L/R}})}}^2) $ of
\begin{equation}
 {\mathcal{A}}(t)
=
\sum_{{\mathbf{k}}, k_x^{\prime}}
a_{{\text{L}},{\mathbf{k}}}(t)  
a^{\dagger }_{{\text{R}},{\mathbf{k}}^{\prime}}(t).
\label{eq:Appendix02}
\end{equation}
The number operator of magnons in the left insulator is
$  N_{\text{L}} =  \sum_{{\mathbf{q}}} 
a^{\dagger }_{{\text{L}}, {\mathbf{q}}} a_{{\text{L}}, {\mathbf{q}}} $.
Through the Heisenberg equation of motion,
we define the operator for the magnon current 
across the junction interface
$ I_{\text{m}} (t)$, from the right to the left insulator,
as
\begin{equation}
I_{\text{m}} (t)
\coloneqq \frac{1}{i \hbar}[N_{\text{L}}, {\mathcal{H}}_{\text{ex}}].
\label{eq:MagnonCurrent}
\end{equation}
Then, the magnon spin current,
i.e., the spin current carried by magnons,
across the junction interface
$I (t) \coloneqq  g \mu _{\text{B}} I_{\text{m}} (t)   $~\cite{magnon2,magnonWF,KNmagnonNoiseJunction}
is given as 
\begin{equation}
I (t)
= - i  g \mu _{\text{B}} \frac{J_{\text{ex}}}{\hbar} 
\sum_{{\mathbf{k}}, k_x^{\prime}}
a_{{\text{L}},{\mathbf{k}}}(t)  
a^{\dagger }_{{\text{R}},{\mathbf{k}}^{\prime}}(t) 
+ {\text{H. c.}},
\label{eq:MagnonSpinCurrent}
\end{equation}
which results in Eqs.~\eqref{eq:Appendix0}
and~\eqref{eq:Appendix02}.

Using the contour-ordered Green's function~\cite{haug,kita,tatara}
we perform the perturbation calculation 
in $ \Gamma_{0({\text{L/R}})}$
based on the Schwinger-Keldysh formalism~\cite{Schwinger,Schwinger2,Keldysh},
and up to $ {\Gamma_{0({\text{L/R}})}}^2$,
which provides
\begin{equation}
\langle {\mathcal{A}}(t) \rangle
=\frac{1}{\hbar^2} 
\int_{{\text{c}}} d \tau_1
\int_{{\text{c}}} d \tau_2
\Gamma_{\text{L}}(\tau_1)
\Gamma_{\text{R}}(\tau_2)  
G_{{\text{L}}, {\mathbf{k}}=0}(\tau, \tau_1)
G_{{\text{R}}, {\mathbf{k}}=0}(\tau_2, \tau)|_{\tau \rightarrow t},
\label{eq:Appendix1}
\end{equation}
where
the applied laser magnetic field excites the zero mode of magnons.
Taking the Schwinger-Keldysh closed time path~\cite{rammer}
$ {\text{c}}={\text{c}}_{\rightarrow}+{\text{c}}_{\leftarrow}  $,
\begin{equation}
\int_{{\text{c}}={\text{c}}_{\rightarrow}+{\text{c}}_{\leftarrow}} d \tau_1
\int_{{\text{c}}={\text{c}}_{\rightarrow}+{\text{c}}_{\leftarrow}} d \tau_2
=
\int_{{\text{c}}_{\rightarrow}} d \tau_1
\int_{{\text{c}}_{\rightarrow}} d \tau_2
+
\int_{{\text{c}}_{\rightarrow}} d \tau_1
\int_{{\text{c}}_{\leftarrow}} d \tau_2 
+
\int_{{\text{c}}_{\leftarrow}} d \tau_1
\int_{{\text{c}}_{\rightarrow}} d \tau_2
+
\int_{{\text{c}}_{\leftarrow}} d \tau_1
\int_{{\text{c}}_{\leftarrow}} d \tau_2,
\label{eq:Appendix2}
\end{equation}
and
using the Langreth theorem~\cite{LangrethTheorem,tatara,haug},
it becomes
\begin{align}
\langle {\mathcal{A}}(t) \rangle
=& 
\frac{1}{\hbar^2} 
\int_{-\infty}^{\infty}dt_1
\int_{-\infty}^{\infty}dt_2
\Gamma_{\text{L}}(t_1)
\Gamma_{\text{R}}(t_2)  
\label{eq:Appendix3} \\
&\cdot
[G^{\text{t}}_{{\text{L}},  {\mathbf{k}}=0}(t, t_1)
G^{\text{t}}_{{\text{R}},  {\mathbf{k}}=0}(t_2, t)  
-G^{\text{t}}_{{\text{L}},  {\mathbf{k}}=0}(t, t_1)
G^{\text{>}}_{{\text{R}},  {\mathbf{k}}=0}(t_2, t)
-G^{\text{<}}_{{\text{L}},  {\mathbf{k}}=0}(t, t_1)
G^{\text{t}}_{{\text{R}},  {\mathbf{k}}=0}(t_2, t) 
+G^{\text{<}}_{{\text{L}},  {\mathbf{k}}=0}(t, t_1)
G^{\text{>}}_{{\text{R}},  {\mathbf{k}}=0}(t_2, t)]. 
\nonumber 
\end{align}
Since~\cite{haug,kita,tatara}
$ G^{\text{t}}_{{\mathbf{k}}}(t, t_1)  
= G^{\text{r}}_{{\mathbf{k}}}(t, t_1) 
+ G^{<}_{{\mathbf{k}}}(t, t_1) $
and
$ G^{<}_{{\mathbf{k}}}(t, t_1)  
- G^{>}_{{\mathbf{k}}}(t, t_1) 
= G^{\text{a}}_{{\mathbf{k}}}(t, t_1) 
- G^{\text{r}}_{{\mathbf{k}}}(t, t_1) $,
it reduces to
\begin{equation}
\langle {\mathcal{A}}(t) \rangle
= 
\frac{1}{\hbar^2} 
\int_{-\infty}^{\infty}dt_1
\int_{-\infty}^{\infty}dt_2
\Gamma_{\text{L}}(t_1)
\Gamma_{\text{R}}(t_2)  
G^{\text{r}}_{{\text{L}}, {\mathbf{k}}=0}(t, t_1)
G^{\text{a}}_{{\text{R}}, {\mathbf{k}}=0}(t_2, t),
\label{eq:Appendix4}
\end{equation}
where
\begin{equation}
\Gamma_{\text{L}}(t_1)
\Gamma_{\text{R}}(t_2)  
=
\Gamma_{0(\text{L})} \Gamma_{0(\text{R})}
[{\text{e}}^{i\Omega(t_1+t_2)}
+{\text{e}}^{i\Omega(t_1-t_2)}
+{\text{e}}^{i\Omega(-t_1+t_2)}
+{\text{e}}^{i\Omega(-t_1-t_2)}].
\label{eq:Appendix5}
\end{equation}
The Fourier transform of the Green's functions results in
\begin{align}
\langle {\mathcal{A}}(t) \rangle
=&
\Gamma_{0(\text{L})} 
\Gamma_{0(\text{R})}
\int_{-\infty}^{\infty}dt_1
\int_{-\infty}^{\infty}dt_2
\int \frac{d\omega_1}{2 \pi}
\int \frac{d\omega_2}{2 \pi}
G^{\text{r}}_{{\text{L}},  {\mathbf{k}}=0, \omega_1}
G^{\text{a}}_{{\text{R}},  {\mathbf{k}}=0, \omega_2}
{\text{e}}^{i(-\omega_1+\omega_2)t}  
\nonumber \\
&{\cdot}[
 {\text{e}}^{i(\omega_1+\Omega)t_1}{\text{e}}^{i(-\omega_2+\Omega)t_2}
+{\text{e}}^{i(\omega_1+\Omega)t_1}{\text{e}}^{i(-\omega_2-\Omega)t_2}
+{\text{e}}^{i(\omega_1-\Omega)t_1}{\text{e}}^{i(-\omega_2+\Omega)t_2}
+{\text{e}}^{i(\omega_1-\Omega)t_1}{\text{e}}^{i(-\omega_2-\Omega)t_2}].
\label{eq:Appendix6}
\end{align}
Performing the time integral 
and the resulting Dirac delta function for the frequency,
we obtain
\begin{align}
 \langle {\mathcal{A}}(t) \rangle
=
\Gamma_{0(\text{L})}
\Gamma_{0(\text{R})}
[&
G^{\text{r}}_{{\text{L}}, {\mathbf{k}}=0, -\Omega}
G^{\text{a}}_{{\text{R}}, {\mathbf{k}}=0, \Omega}
{\text{e}}^{2i\Omega t} 
+
G^{\text{r}}_{{\text{L}}, {\mathbf{k}}=0, \Omega}
G^{\text{a}}_{{\text{R}}, {\mathbf{k}}=0, -\Omega}
{\text{e}}^{-2i\Omega t}   
\nonumber  \\
&+
(G^{\text{r}}_{{\text{L}}, {\mathbf{k}}=0, -\Omega}
G^{\text{a}}_{{\text{R}},  {\mathbf{k}}=0, -\Omega} 
+
G^{\text{r}}_{{\text{L}},  {\mathbf{k}}=0, \Omega}
G^{\text{a}}_{{\text{R}},  {\mathbf{k}}=0, \Omega})].
\label{eq:Appendix7}
\end{align}
This is summarized as follows:
\begin{equation}
 \langle {\mathcal{A}}(t) \rangle
=
\Gamma_{0(\text{L})}
\Gamma_{0(\text{R})}
\sum_{n=0,\pm2} 
A_n {\text{e}}^{ni\Omega t},
\label{eq:Appendix8}
\end{equation}
with the coefficients
\begin{subequations}
\begin{align}
A_2
&=
G^{\text{r}}_{{\text{L}},  {\mathbf{k}}=0, -\Omega}
G^{\text{a}}_{{\text{R}},  {\mathbf{k}}=0, \Omega}, 
\label{eq:AppendixA2}  \\
A_{-2}
&=
G^{\text{r}}_{{\text{L}},  {\mathbf{k}}=0, \Omega}
G^{\text{a}}_{{\text{R}},  {\mathbf{k}}=0, -\Omega}, 
\label{eq:AppendixA-2} \\
A_0
&=
G^{\text{r}}_{{\text{L}},  {\mathbf{k}}=0, -\Omega}
G^{\text{a}}_{{\text{R}},  {\mathbf{k}}=0, -\Omega}
+
G^{\text{r}}_{{\text{L}},  {\mathbf{k}}=0, \Omega}
G^{\text{a}}_{{\text{R}},  {\mathbf{k}}=0, \Omega}.
\label{eq:AppendixA0}
\end{align}
\end{subequations}
The coefficients are described
in terms of the magnon lifetime and the energy gap for the zero mode
as
\begin{subequations}
\begin{align}
A_2
&\eqqcolon
A_{\Omega}
=
-\frac{[\hbar \Omega+\Delta_{\text{L}}+i\hbar/(2\tau_{\text{L}}) ][\hbar \Omega - \Delta_{\text{R}}+i\hbar/(2\tau_{\text{R}})]}{[(\hbar \Omega + \Delta_{\text{L}})^2+(\hbar/2\tau_{\text{L}})^2][(\hbar \Omega - \Delta_{\text{R}})^2+(\hbar/2\tau_{\text{R}})^2]}, 
\label{eq:Appendix9a} \\
A_{-2}
&=
A_{-\Omega},  
\label{eq:Appendix9b} \\   
A_0
&=
\frac{[\hbar \Omega+\Delta_{\text{L}}+i\hbar/(2\tau_{\text{L}}) ]
[\hbar \Omega + \Delta_{\text{R}}-i\hbar/(2\tau_{\text{R}})]}
{[(\hbar \Omega + \Delta_{\text{L}})^2+(\hbar/2\tau_{\text{L}})^2]
[(\hbar \Omega + \Delta_{\text{R}})^2+(\hbar/2\tau_{\text{R}})^2]}
+
\frac{[-\hbar \Omega+\Delta_{\text{L}}+i\hbar/(2\tau_{\text{L}}) ]
[-\hbar \Omega + \Delta_{\text{R}}-i\hbar/(2\tau_{\text{R}})]}
{[(\hbar \Omega - \Delta_{\text{L}})^2+(\hbar/2\tau_{\text{L}})^2]
[(\hbar \Omega - \Delta_{\text{R}})^2+(\hbar/2\tau_{\text{R}})^2]}.
\label{eq:Appendixpc}
\end{align}
\end{subequations}

\end{widetext}

\bibliography{PumpingRef}
\end{document}